\documentclass[twocolumn]{aastex631}
\usepackage{amsmath}
\usepackage{xspace}
\usepackage{soul}
\newlength{\wdth}

\newcommand{\xmm}{{\it XMM-Newton}\xspace}

\newcommand{\cgs}{{\rm erg s$^{-1}$ cm$^{-2}$}\xspace}

\newcommand{\target}{{eRO-QPE2}\xspace}
\newcommand{\rx}{{RX~J1301.9+2747}\xspace}

\newcommand\soutpars[1]{\let\helpcmd\sout\parhelp#1\par\relax\relax}
\long\def\parhelp#1\par#2\relax{%
  \helpcmd{#1}\ifx\relax#2\else\par\parhelp#2\relax\fi%
}

\expandafter\ifx\csname natexlab\endcsname\relax\fi
\providecommand{\url}[1]{\href{#1}{#1}}
\providecommand{\dodoi}[1]{doi:~\href{http://doi.org/#1}{\nolinkurl{#1}}}
\providecommand{\doeprint}[1]{\href{http://ascl.net/#1}{\nolinkurl{http://ascl.net/#1}}}
\providecommand{\doarXiv}[1]{\href{https://arxiv.org/abs/#1}{\nolinkurl{https://arxiv.org/abs/#1}}}

\shorttitle{eRO-QPE2's highly stable QPEs}
\shortauthors{Pasham et al.}

\graphicspath{{./}{}}

\begin{document}

\title{Alive and Strongly Kicking: Stable X-ray Quasi-Periodic Eruptions from eRO-QPE2 over 3.5 Years}

\author[0000-0003-1386-7861]{Dheeraj Pasham}
\affiliation{MIT Kavli Institute for Astrophysics and Space Research \\
		Cambridge, MA 02139, USA}





\author[0009-0004-5838-1886]{Shubham Kejriwal}
\affiliation{Department of Physics, National University of Singapore, Singapore 117551}

\author[0000-0003-3765-6401]{E.~R.~Coughlin}
\affiliation{Department of Physics, Syracuse University, Syracuse, NY 13210, USA}

\author[0000-0002-9209-5355]{Vojt{\v e}ch Witzany}
\affiliation{Institute of Theoretical Physics, Charles University, \\ 
V Holešovičkách 2,
180 00 Prague 8, 
Czechia}

\author[0000-0001-5242-8269]{Alvin J. K. Chua}
\affiliation{Department of Physics, National University of Singapore, Singapore 117551}
\affiliation{Department of Mathematics, National University of Singapore, Singapore 119076}

\author[0000-0001-6450-1187]{Michal Zaja\v{c}ek}
\affiliation{Department of Theoretical Physics and Astrophysics, Faculty of Science, Masaryk University,\\ Kotl\'a\v{r}sk\'a 2, 611 37 Brno, Czech Republic}

\author[0000-0002-0786-7307]{T. Wevers}
\affiliation{Space Telescope Science Institute, 3700 San Martin Drive, Baltimore, MD 21218, USA}

\author{Yukta Ajay}
\affiliation{Department of Physics and Astronomy, Johns Hopkins University, 3400 N. Charles Street, Baltimore, MD 21218, USA}
    
\begin{abstract}
Quasi-periodic eruptions (QPEs) are recurring bursts of soft X-rays from the nuclei of galaxies. Their physical origin is currently a subject of debate, with models typically invoking an orbiter around a massive black hole or disk instabilities. Here we present and analyze the temporal and spectral evolution of the QPE source eRO-QPE2 over 3.5 years. We find that eRO-QPE2 1) is remarkably stable over the entire 3.5-year temporal baseline in its eruption peak luminosity, eruption temperature, quiescent temperature, and quiescent luminosity, 2) has a stable mean eruption recurrence time of 2.35 hours, with  \textbf{marginal ($\sim$2$\sigma$) evidence for a $0.1$ hour reduction over the 3.5 yr period}, and 3) has a long-short variation in its recurrence time in August 2020, but this pattern is absent from all subsequent observations. The stability of its peak eruption luminosity and that of the quiescent state are notably dissimilar from three previously tracked QPEs (GSN069, eRO-QPE1, eRO-QPE3), which show declines in eruption and quiescent flux over comparable temporal baselines. This stability is even more pronounced in eRO-QPE2 due to its 2.4 hour average recurrence time compared to GSN-069's 9 hour, eRO-QPE1's 16 hour, and eRO-QPE3's 20 hour recurrence times, i.e., this system has undergone 4-8 times more cycles than these other systems over the 3.5 years of observations. We discuss the implications of these observations within the context of some proposed extreme mass ratio inspiral (EMRI) models. 

\end{abstract}


\section{Introduction}\label{sec:intro}
Quasi-periodic eruptions (QPEs) are recurring bursts of soft X-rays (0.2-3.0 keV) that are spatially coincident with the centers of nearby (redshift, $z\lesssim$0.1) galaxies \citep{gsn069,rxqpes}. There are presently eight  systems in published literature with confirmed QPEs: GSN~069 \citep{gsn069}, \rx \citep{rxqpes}, eRO-QPE1, eRO-QPE2 \citep{arcodia21}, eRO-QPE3, eRO-QPE4 \citep{qpe34}, AT2019qiz \citep{2024arXiv240902181N}, and SwJ023017.0+283603 \citep{evans23, guolo24}. These have recurrence times (i.e., the amount of time between successive eruptions) ranging from 2.4 hours to 22 days, with a dispersion in arrival time of eruptions of up to $\sim 30\%$ \citep{miniutti23a, barelykick, joheen}. In general, their X-ray spectra during quiescence, i.e., between the eruptions, can be fit with a disk blackbody with a temperature of a few tens of eV \citep[e.g.,][]{gsn069,arcodia21}. During the eruptions an additional single-temperature blackbody (0.1-0.25 keV) is necessary to explain the data, and thus the presence of a warmer thermal component is generally correlated with the X-ray flux \citep[e.g.,][]{gsn069}. It has been suggested that the quiescent emission tracks an underlying disk, perhaps formed from a relatively recent tidal disruption event (TDE) \citep[e.g.,][]{rees88, gezari21}. In addition, the host galaxies of QPEs and TDEs have a number of shared preferences, including an overrepresentation of post-starburst galaxies \citep{French16, Graur18, Wevers22} and an overrepresentation of gas-rich environments with recently faded active galactic nuclei \citep{Wevers24a, Wevers24b}. A clear direct connection between an optically selected TDE and an X-ray QPE has only recently been established \citep{2024arXiv240902181N}. 

Alongside other repeating phenomena such as quasi-periodic outflows (QPOuts; \citealt{qpouts}) and stable soft X-ray quasi-periodic oscillations (QPOs; \citealt{shubham, rejqpo, 14liqpo}), these repeating extragalactic nuclear transients (RENTs) could represent electromagnetic counterparts of extreme mass ratio inspirals (EMRIs), which contain a massive black hole and an orbiting companion (a star or another compact object) that is substantially less massive \citep{krolikmodel,itaimodel1, kingqpemodel, repeated_emris_model, itaimodel2, petra, alessiamodel, 2022A&A...661A..55Z, xianmodel, 2024arXiv240207177W, 2023MNRAS.523L..26K}. The alternative hypothesis is that these regular modulations could be triggered by instabilities operating in the inner regions of the accretion flow \citep{marzenamodel, kaurmodel, 2023Ap&SS.368....8C, panmodel, Raj:2021b}. The latter set of models has been disfavored -- at least in some cases -- because the periods appear to be uncorrelated with the black hole's mass (see bottom right panel of Fig. 5 of \citealt{guolo24}), and the shape of the eruption profiles are inverted with respect to what is predicted from the radiation pressure instability (e.g., compare Figs. 1 and 2 of \citealt{arcodia21} with Fig. 7-10 of \citealt{marzenamodel} and Fig. 3 of \citealt{Raj:2021b}). We stress, however, that at present we cannot rule out instability models.


Using EMRI population models, some works have argued that the most favorable orbital frequency (at the present epoch) for enabling future detection with LISA is 0.5$\pm$0.2 mHz, or an orbital period on the order of $\sim 1$ hour \citep{shubham}. With a mean period of 2.4 hours \citep{arcodia21}, eRO-QPE2 (redshift $z=0.0175$; \citealt{arcodia21}) is thus an especially exciting target, as it may represent a promising candidate for multi-messenger study in the coming age of space-based gravitational-wave observatories \citep{2022A&A...661A..55Z}. 

In this work, we studied the long-term evolution of \target using \xmm data taken between August 2020 to February 2024, i.e., a temporal baseline of 1277 days or 3.5 years. Our main finding is that, unlike the three previously tracked QPE systems GSN~069 \citep{miniutti23a}, eRO-QPE1 \citep{barelykick, joheen}, and eRO-QPE3 \citep{qpe34}, eRO-QPE2 has remained stable in its eruption strength, average time between eruptions, and quiescent luminosity (section \ref{sec: results}). The median (standard deviation) time between the 9 eruptions seen in Aug 2020 was 2.42 (0.09) hours. The values combining the data taken in December 2023 and February 2024 is 2.33 (0.06) hours.  Although not statistically significant, the small change of 0.09 hours over $\approx$3 years could represent the orbital decay of the putative EMRI. This would however be too fast for a vacuum EMRI (section \ref{sec:shubham}).

\section{Data Reduction and Analysis}
\label{sec:data}
Within the context of this work, we use a standard $\Lambda$CDM cosmology with parameters H$_{0}$ = 67.4 km s$^{-1}$ Mpc$^{-1}$, $\Omega_{\rm m}$ = 0.315 and $\Omega_{\rm \Lambda}$ = 1 - $\Omega_{\rm m}$ = 0.685 \citep{2020A&A...641A...6P}. Using the Cosmology calculator of \cite{2006PASP..118.1711W}, \target's  luminosity distance is estimated to be 78.9 Mpcs.

\subsection{\xmm Data Reduction and Analysis}\label{sec:xmmdata}
\xmm's European Photon Imaging Camera (EPIC; pn: \citealt{epicpn}, MOS: \citealt{epicmos}) observed \target on six occasions between 6 August 2020 and 4 February 2024. One of the observation is not public and we did not include it in our work. The two most recent observations (obsIDs: 0932590101/XMM\#5, 0932590201/XMM\#4) were part of an approved guest observer program (PI: Wevers T.) and we include them in this work along with the 3 publicly available datasets (obsIDs: 0872390101/XMM\#1, 0893810501/XMM\#2, 0883770201/XMM\#3). We used data from \xmm's European Photon Counting Camera (EPIC) pn and MOS in this work. Combined EPIC (pn+MOS) data was used for light curve analysis to improve the statistics in individual eruptions. However, for energy spectral analysis we exclude MOS data because of their deteriorated response below $\sim$1 keV, e.g., see \url{https://xmmweb.esac.esa.int/docs/documents/CAL-TN-0018.pdf}. We used \xmm's software XMMSAS version 19.1.0 with the latest calibration data for analysis. 

\begin{figure*}[ht]
    \centering
    \includegraphics[width=0.9\textwidth]{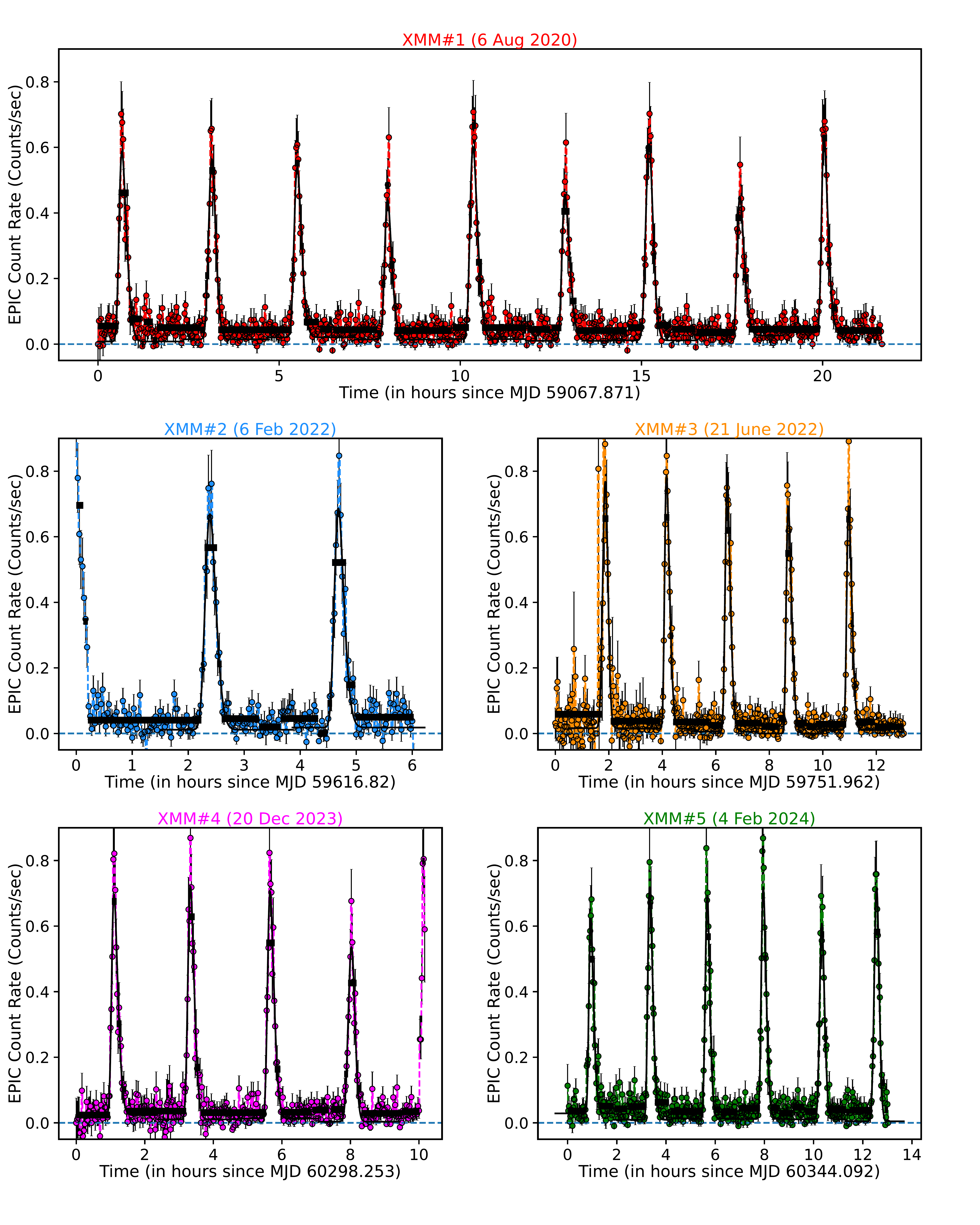}
    \caption{ {\bf 0.25-2.5 keV \xmm/EPIC (pn+MOS) X-ray light curves of \target.} The time bin size in each case is 100 s and the observation dates are indicated at the top of each panel. The thick black horizontal lines are the optimal time bins derived from the Bayesian blocks algorithm of \cite{blocks2}. The solid curves are the best-fit skewed-Gaussian model fits.}
    \label{fig:fig1}
\end{figure*}

First we downloaded the five data from \xmm's science archive accessible at \url{https://www.cosmos.esa.int/web/xmm-newton/xsa}. We reduced the raw Observation Data Files (ODFs) using the standard procedures outlined in these data analysis threads: \url{https://www.cosmos.esa.int/web/xmm-newton/sas-threads}. Then we extracted source events separately from pn and MOS detectors. For this we used a circular aperture centered on coordinates (RA, Dec) = (02:34:48.97, -44:19:31.65) with a radius of 25$^{\prime\prime}$. For pn (MOS) we screened out events with PATTERN greater than 4 (12). We used 0.25-2.5 keV bandpass where the source is detected above the background. A nearby circular regions with a radii of 50$^{\prime\prime}$ and free of point sources was chosen to compute the background. For each obsID we inspected the background light curves manually and excluded epochs dominated by flares. We combined the instrumental good time intervals (GTIs) with those excluding the background flares to obtain a final clean GTI for each obsID. The resulting X-ray light curves are shown in Fig. \ref{fig:fig1}.

From each obsID we extracted two spectra using only pn data: one covering the epochs of the eruptions and another using events during the quiescence. The spectra were binned using XMMSAS software's {\it specgroup} task with {\it mincounts}=1 and {\it oversample}=3. C-stat was used for fitting. For all spectra, the MilkyWay Hydrogen column of {\it tbabs} was derived from HI maps using the HEASARC online tool \url{https://heasarc.gsfc.nasa.gov/cgi-bin/Tools/w3nh/w3nh.pl} and was fixed to a value of 1.6$\times$10$^{20}$ cm$^{-2}$. The best-fitting model parameters are show in Fig. \ref{fig:fig3}. {\it XSPEC} \citep{xspec} ready X-ray spectra along with the background spectra and relevant response files can be found here: \url{https://zenodo.org/records/13140806}.

\section{Results}\label{sec: results}
\begin{figure*}[t]
    \centering
    \includegraphics[width=1.05\textwidth]{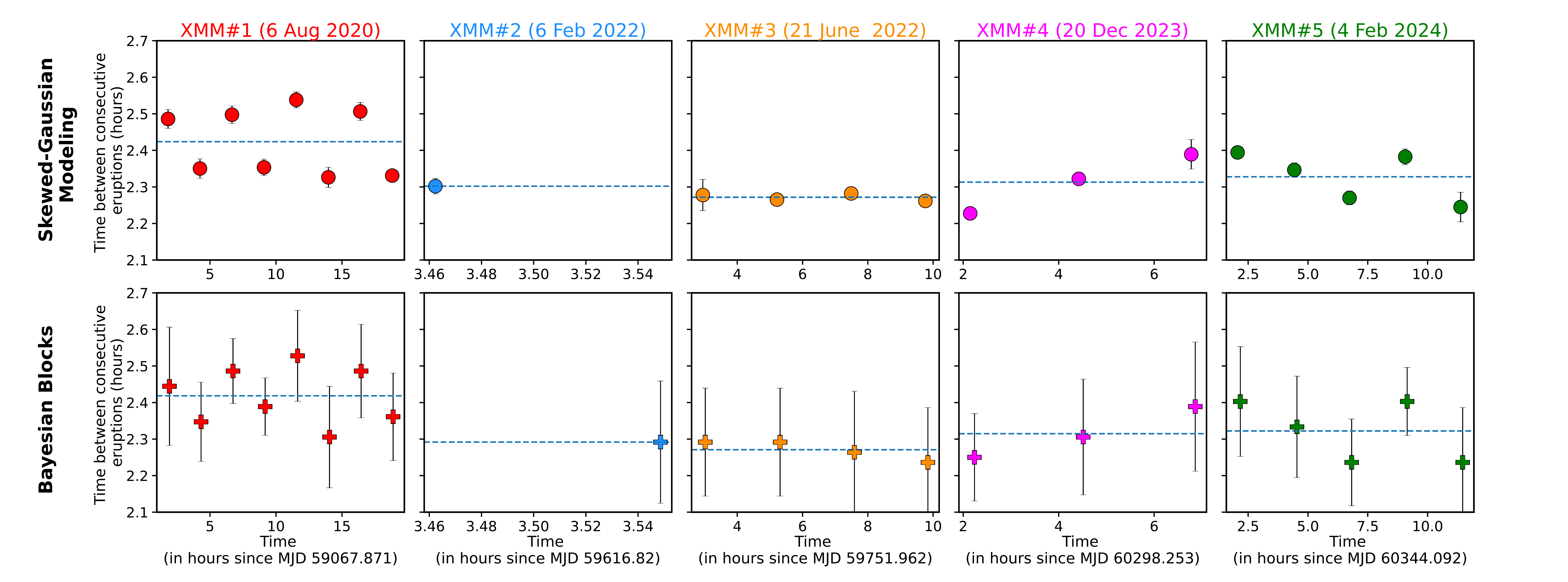}
    \caption{{\bf Evolution of time between eruptions with time for all \xmm datasets.} The horizontal dashed blue lines represent the mean value in each case. The y-scale is the same in all panels (2.1 to 2.7 hours). Panels in the same column share their x-axes. Bayesian blocks is a model independent way of estimating the peak times while in the top row we show peak times estimated by modeling the eruptions with skewed Gaussians. The errorbars in the case of Bayesian blocks represent the size of the block while it represents the statistical uncertainty for skewed Gaussian modeling.}
    \label{fig:fig2}
\end{figure*}

Fig. \ref{fig:fig1} shows the five \xmm/EPIC (pn+MOS) 0.25-2.5 keV X-ray light curves of \target. We applied the Bayesian blocks algorithm \citep{blocks2} to estimate the peaks of the individual eruptions in a model-independent manner (thick, black horizontal lines in Fig. \ref{fig:fig1}). These values are shown in the bottom row of Fig. \ref{fig:fig2}. We then fit the eruptions with a skewed-Gaussian model that allows us to estimate the peaks to a much higher precision--typical 1$\sigma$ uncertainty of $<$100 secs (top row of Fig. \ref{fig:fig2}).  

In order to reliably estimate the median time between eruptions we need to sample several of them. The best dataset we have is XMM\#1 with 9 eruptions, followed by XMM\#5, XMM\#3 and XMM\#4 with 6, 5 and 4 full eruptions, respectively. The mean time (standard deviation) between subsequent eruptions during XMM\#1, XMM\#3, XMM\#4, and XMM\#5 was 2.42 (0.09), 2.27 (0.01), 2.32 (0.07), and 2.35 (0.06), respectively. If we exclude XMM\#2, which had only two poorly sampled eruptions, then we can see that the mean time between eruptions during mid-2022 and late 2023--early 2024 has  perhaps decreased by about 0.1 hours compared to the eruptions on 6 August 2020.

\subsection{Quantifying any potential long-term trend}\label{sec:linmix}
To test for a trend in the period with the time, we used the {\tt linmix} package \citep{2007ApJ...665.1489K}. {\tt linmix} is a Bayesian framework for linear regression that finds best-fit linear regression parameters (slope and intercept) with uncertainties and a correlation index (Pearson), by taking into account errors in both x- and y-values. The documentation and more details on {\tt linmix} implementation can be found at this {\tt github} repository \url{https://github.com/jmeyers314/linmix?tab=readme-ov-file}.

\begin{figure*}[ht]
    \centering
    \includegraphics[width=0.95\textwidth]{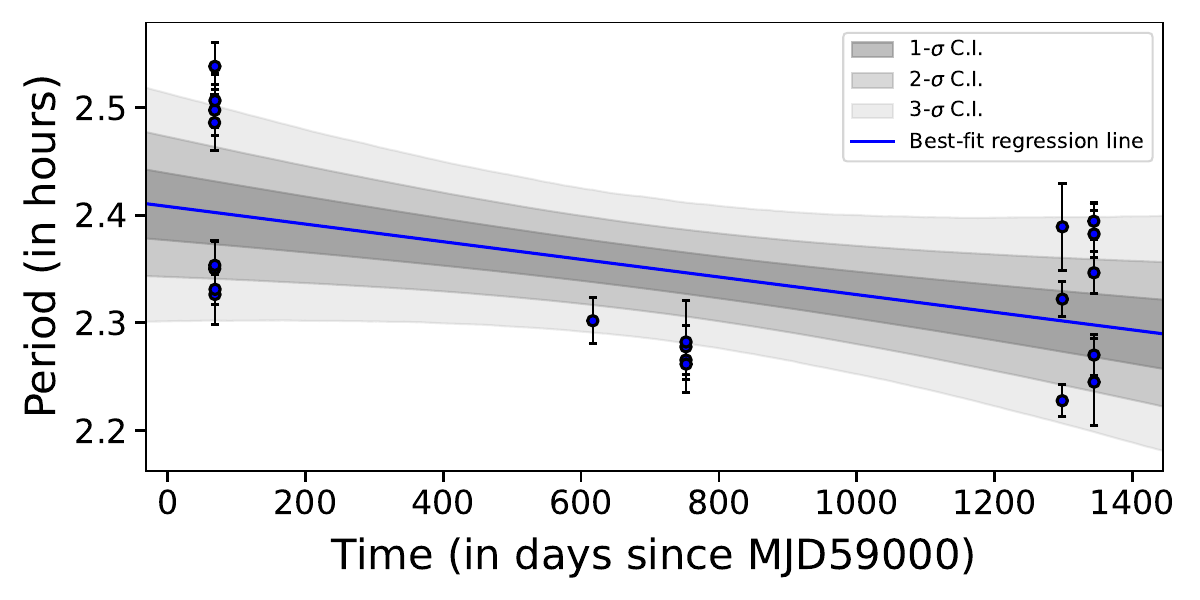}
    \caption{ \textbf{Best-fit regression fit to the time between eruptions vs time using the {\tt linmix} linear regression framework.}  The shaded grey regions give 1$\sigma$, 2$\sigma$ and 3$\sigma$ confidence intervals (C.I.) of the computed best-fit line. Both x and y errors have been included. The x-errors are not visible since the error bars are smaller than the marker size. The evidence for a linear decay of about 0.1 hr over 3+ years is marginal (see Fig.~\ref{fig:fig6}). }
    \label{fig:fig5}
\end{figure*}

\begin{figure*}[ht]
    \centering
    \includegraphics[width=0.95\textwidth]{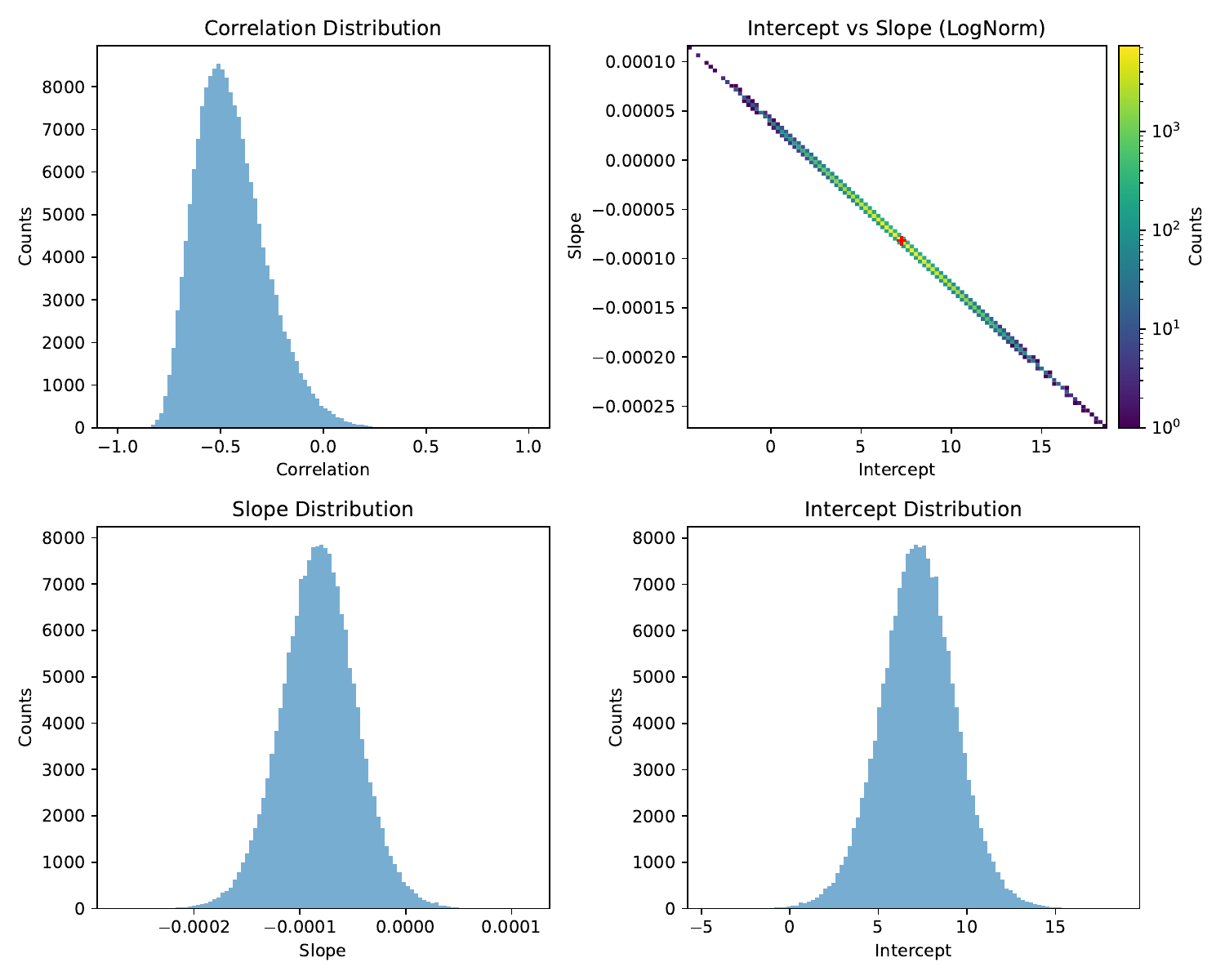}
    \caption{ {\bf Results from the {\tt linmix} regression analysis of time between eruptions vs time.} (top left) Distribution of the correlation (Pearson index); (top right) Log-normal distribution of the slope and intercept values, the red cross marker gives the location of best-fit regression values in the parameter space (marker size not to be scaled with associated errors); Distribution of the fit values of (bottom left) slope and (bottom right) intercept.}
    \label{fig:fig6}
\end{figure*}

We carried out the regression analysis using the {\it LinMix} function by fitting the data points using 2 Gaussians (K = 2) and instantiating 50 Monte Carlo Markov Chains ({\it nchains} = 50) for 10000 iterations. The first 30\% of the fit values were discarded since this fraction corresponds to the ``burn-in'' phase of the MCMC sampling. We calculate the best-fit regression parameters by finding the median of the parameter distributions for slope and intercept, as a median estimate is less-sensitive to outliers in the data. Results of the regression analysis are shown in Fig. \ref{fig:fig5}  and \ref{fig:fig6} and the best-fit parameters are slope=(-8.2$\pm$3.6)$\times$10$^{-5}$ hours/day and intercept=(7.3$\pm$2.1) hours. The Pearson correlation index takes values from -1 to 1, where an index $>$0 suggests a positive correlation, values close to 0 suggest no (or weak) linear correlation, and values $<$0 point towards a negative (or inverse) correlation. 

The above analysis suggests that the evidence for a decreasing trend is only marginal at about 2$\sigma$ (see Fig.~\ref{fig:fig6}).

\subsection{Long-Term Spectral Evolution}\label{sec:specevol}
Next, we studied the evolution of the average eruption and quiescence spectrum over the 3.5 year period. Similar to previous studies of QPEs \citep[e.g.,][]{gsn069, arcodia21}, we fit the eruption spectra with a single temperature blackbody {\it XSPEC} \citep{xspec}: {\it tbabs*ztbabs*zashift(bbody)}, and the quiescent spectra with a single disk blackbody: {\it tbabs*ztbabs*zashift(diskbb)}. The evolution of the resulting best-fit model parameters is shown in Fig. \ref{fig:fig3}. \textbf{We then studied the stability of the individual eruption peaks and widths which are shown in Fig.~\ref{fig:peakvstime} and \ref{fig:widthvstime}. Based on this we conclude that the \target's eruptions have been stable over the past 3.5 years.} 


\section{Comparison to long-term evolution of other QPEs}\label{sec:comp}
Four QPE sources, GSN~069, eRO-QPE1, eRO-QPE3 and eRO-QPE4 have been tracked over multiple years \citep{miniutti23a, barelykick, joheen, qpe34}. In the first three cases, the average strength of eruptions/peak eruption flux gradually decreased over a few years timescale  (see Figs. 1 and 2 of \citealt{miniutti23a}, Fig. 2 of \citealt{barelykick} and Fig. 11 of \citealt{qpe34}). In the case of GSN~069 QPEs shut off over roughly 500 days but turned back on after about two years (see Fig. 4 of \citealt{alivenkicking}). eRO-QPE4's data lacks the signal-to-noise ratio necessary to determine if similar behavior is occurring. Further tracking is necessary to see if eRO-QPE3 and eRO-QPE1's eruptions shutoff and turn back on in a manner similar to GSN~069.

In GSN~069, eRO-QPE1 and eRO-QPE3 there is an apparent declining trend in observed quiescence X-ray luminosity (see Fig. 4 of \citealt{alivenkicking} and Fig. 11 of \citealt{qpe34}, and \citealt{joheen}). In the case of GSN~069 and eRO-QPE3 this represents a factor of $\approx$3 change over 500 days and $\approx$5 decrease over 800 days, respectively. Based on our analysis of eRO-QPE1's most recent \xmm dataset taken in January 2024 (PI: Arcodia) and its early \xmm observations, we estimate a decrease of roughly a factor of 2.5 between July--August 2020 and January 2024 (observed 0.3-1.2 keV quiescent fluxes of (4.7$\pm$0.7)$\times$10$^{-15}$ \cgs and (2.1$\pm$0.3)$\times$10$^{-15}$ \cgs). In summary, GSN~069, eRO-QPE1 and eRO-QPE3 appear to behave the same way over a 3+ years in terms decreasing eruption strength and quiescence luminosity. 

\target's long-term behavior is distinct from all the above QPE systems: \target has been remarkably stable in terms of average eruption luminosity, eruption temperature, quiescence luminosity and temperature (see Fig. \ref{fig:fig2}). 

\section{Basic energetics calculation}\label{sec:energetics}
With a redshift of $z=0.0175$ \citep{arcodia21}, the luminosity per eruption in the 0.2 -- 2.5 keV band is $\simeq 10^{42.2}$ erg s$^{-1}$, which translates to a total integrated luminosity of $\sim 10^{43.3}$ erg s$^{-1}$. If the liberated energy ultimately derives from accretion onto the black hole, then adopting a radiative efficiency of $0.1$ and an eruption duration of $\sim 2$ ks implies an accreted mass per eruption of $\sim 2.2\times 10^{-7} M_{\odot}$. If the object feeding the accretion has a mass comparable to a solar mass, then the total number of eruptions required to completely deplete the mass of the object is $\sim 4.5\times 10^{6}$, suggesting that the lifetime of the system is $\sim 4.5\times 10^{6}\times 2.4\textrm{ hr} \simeq 1200$ yr. If we instead use only the energy in the 0.2 -- 2.5 keV band, the lifetime would be increased by a factor of 10, but regardless the system would be cosmologically short lived. In case the eruptions are associated with a thermally emitting area, a typically lengthscale is $R_{\rm erupt}\sim (L_{\rm erupt}/10^{43.3}\,{\rm erg\,s^{-1}})^{1/2} (kT_{\rm erupt}/200\,{\rm eV})^{-2}\sim 28\,R_{\odot}$ or 133 gravitational radii for $M_{\bullet}=10^5\,M_{\odot}$. Hence, the emission could be associated with an ejected, expanding cloud \citep{alessiamodel} or a compact area of an accretion disk emitting due to circularization shocks with an inspiralling gas \citep{wenbinmodel}. 

\begin{figure}[ht]
    \centering
    \includegraphics[width=0.49\textwidth]{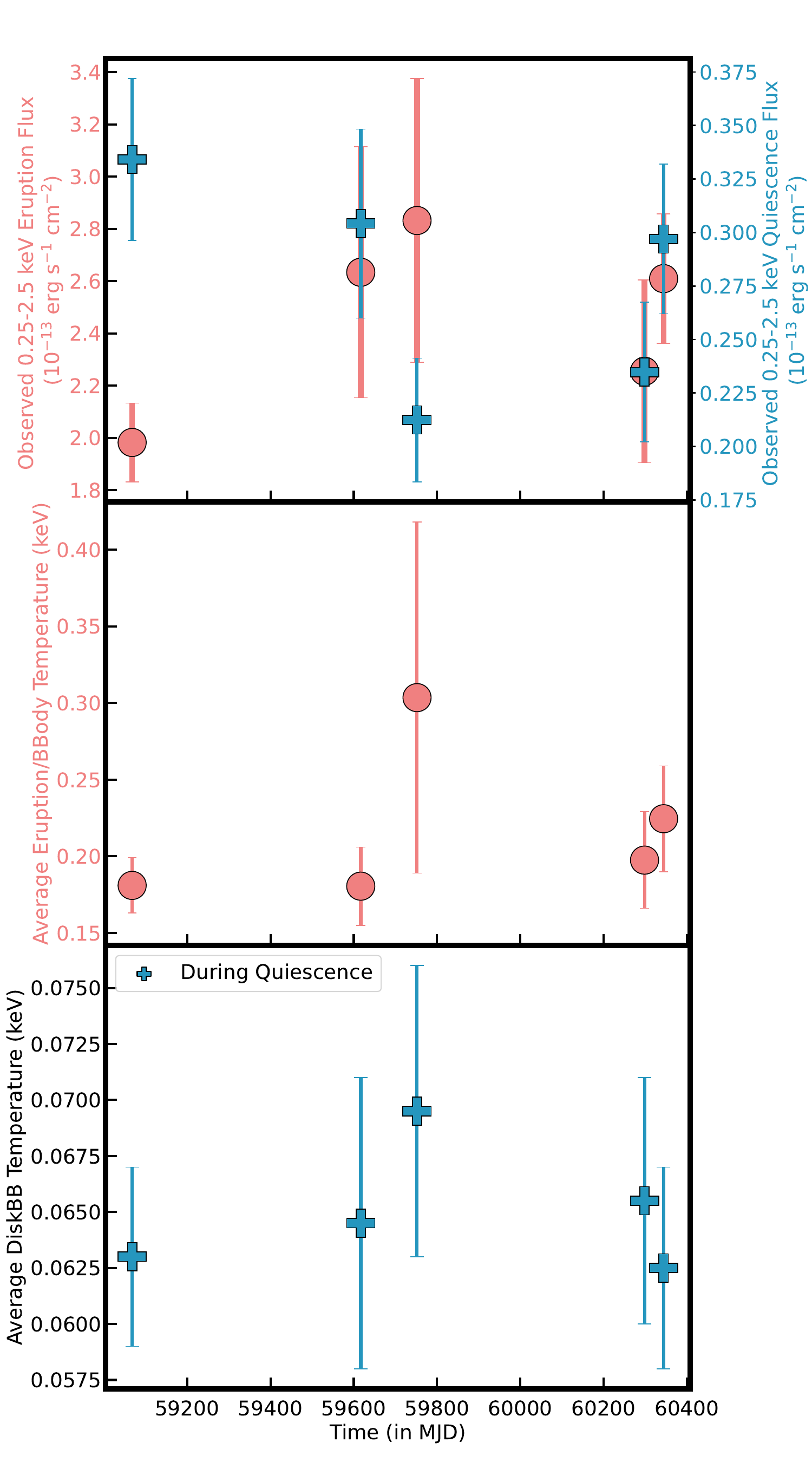}
    \caption{{\bf Long-term evolution of \target's spectra during eruptions and quiescence.} From each \xmm obsID two spectra were derived: one using data during eruptions and one during the quiescence. These spectra were fit with a disk blackbody and a pure blackbody. During the quiescence only disk blackbody was sufficient. All errorbars represent 90\% uncertainties. This data is available at \url{https://zenodo.org/records/11415786}}
    \label{fig:fig3}
\end{figure}

\begin{figure*}[ht]
    \centering
    \includegraphics[width=1.05\textwidth]{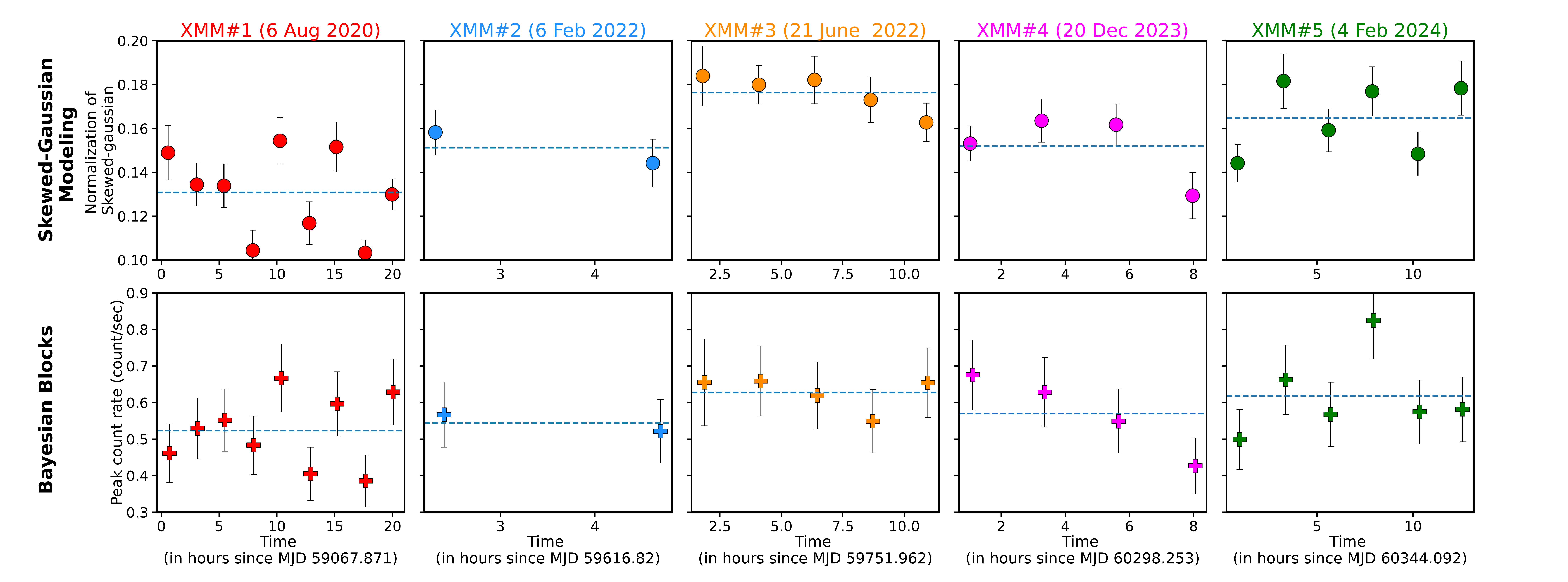}
    \caption{{\bf Same as Fig~\ref{fig:fig2} but here we show the evolution of eruption peaks with time for all \xmm datasets.} }
    \label{fig:peakvstime}
\end{figure*}

\begin{figure*}[ht]
    \centering
    \includegraphics[width=1.05\textwidth]{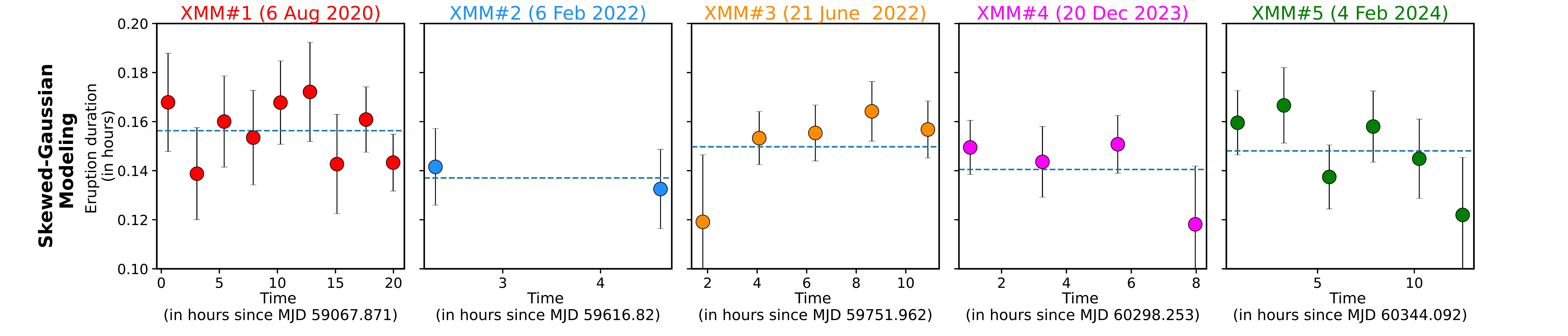}
    \caption{{\bf Same as Fig~\ref{fig:fig2} but here we show the evolution of the eruption duration (width of the skewed Gaussians) with time for all \xmm datasets} }
    \label{fig:widthvstime}
\end{figure*}

A recent set of hydrodynamical simulations by \citet{Yao:2024rtl} showed that a main-sequence star should be stripped of $\sim 10^{-6}-10^{-4} M_{\odot}$ per passage through the disk, significantly more than suggested by the energetics above. Following \citet{itaimodel2} the authors modeled the QPE energy release only as a fraction of the energy of the shock caused by the supersonic star-disk collision, implying that the rest of the matter builds up in the accretion disk. Using this assumption, the authors estimated the lifetime of eRO-QPE2 to mere decades due to the ablation of the star, which also led to a suggestion of a future rise or outburst of the quiescent accretion luminosity due to the matter build-up in the disk. This is in tension with our observation in Fig.~\ref{fig:fig3}, which shows that both the strength of the QPEs and the quiescent emission of the disk is stable on the timescale of years.

\section{Discussion}\label{sec:discussion}
\subsection{On the disappearance of long--short pattern}
\begin{figure}[ht]
    \centering
    \includegraphics[width=0.8\linewidth]{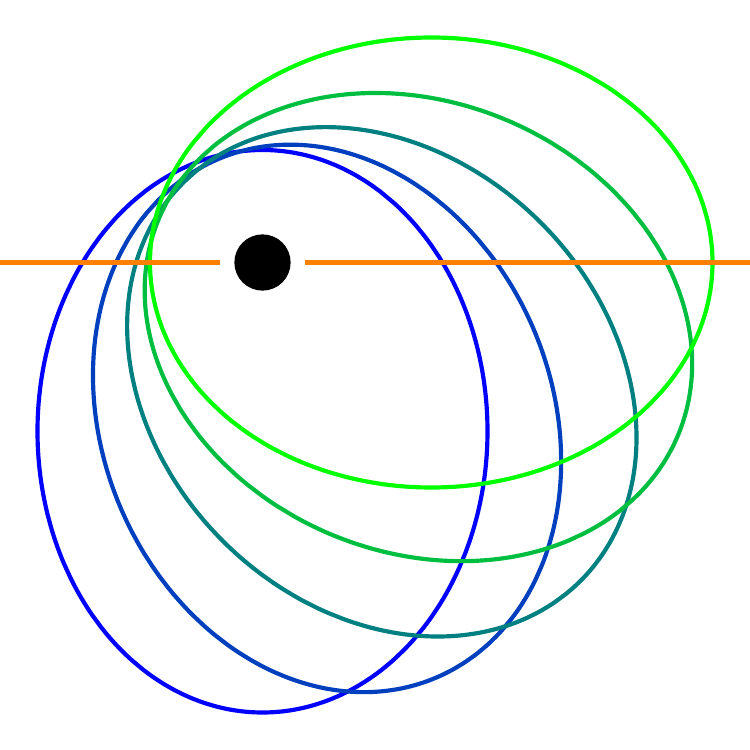}
    \caption{{\bf The potential scenario for the disappearance of the long-short pattern. In 2020 we see the shorter times between eruptions because of the faster pericentre passage, and the longer time due to the apocenter passage (blue curve). Due to precession the orbit shifts to the green ellipse where the passage between intersections becomes the same and the long-short pattern disappears.} }
    \label{fig:precess}
\end{figure}

One of the clear results from our study is the disappearance of the long--short pattern seen in the first \xmm observation (top--left of Fig.~\ref{fig:fig2}). Under the EMRI paradigm the long--short pattern can be explained with an eccentric orbiter interacting with the disk twice per orbit. The setup is illustrated in Figure \ref{fig:precess}. We assume that the orbit is mildly eccentric and that the short intervals $T_{\rm s}$ correspond to the section of the orbit containing the comparatively quick pericentre passage and the long intervals $T_{\rm l}$ to sections containing apocenter passages.  We use the small-eccentricity expansion of the Kepler equation to express the times as
\begin{equation}
\begin{split}
    & T_{\rm s} = \frac{P_{\rm orb}}{2} \left(1 - \frac{4 e}{\pi}\right) + \mathcal{O}(e^3)\,, 
    \\
    & T_{\rm l} = \frac{P_{\rm orb}}{2} \left(1 + \frac{4 e}{\pi}\right) + \mathcal{O}(e^3)\,.
    \end{split}
\end{equation}
From this we obtain $e \approx \pi (T_{\rm l} - T_{\rm s})/(4 P_{\rm orb})$
Using the approximate $P_{\rm orb} \approx 4.8$ hours and $T_{\rm l} - T_{\rm s} \approx 0.15$ hours we get $e \approx 0.025$. The disappearance of the long-short pattern can then be accounted for by relativistic pericenter precession as illustrated in Figure \ref{fig:precess}, which leads to equal times between passages and occurs with a period \citep{robertson1938note}
\begin{equation}
\begin{split}
    T_{\rm prec} &= \frac{2 \pi P_{\rm orb}^{5/3} c^2}{3 (2 G M \pi)^{2/3}} + \mathcal{O}(e^2) 
    \\
    & = 30 \, {\rm days} \left(\frac{M}{10^6 M_\odot}\right)^{-2/3}  \left(\frac{P_{\rm orb}}{5\,\rm hours}\right)^{-5/3}.
\end{split}
\end{equation}
The switch between the equal-passage and long-short transition times occurs twice per a full precession cycle, so we can essentially assume a random pattern to appear in the XMM\#1-\#5 datasets that spread over years. This qualitatively fits our observations. To confirm this scenario, we would require a good mass estimate on the BH in \target to constrain $T_{\rm prec}$ and a series of $\sim 25$-hour observations repeated a few times over the timespan $T_{\rm prec}/2$.

\subsection{Stability of the quiescent emission}
\label{sec:stability-quiescent}
It has been hypothesized that the disc through which the orbiter passes (thus generating the QPEs) can be produced from a TDE \citep{itaimodel2, 2024arXiv240902181N}, and -- if the fallback of debris traces the accretion rate onto the black hole -- the declining amplitude of the quiescent emission seen in other sources \citep{qpe34, alivenkicking} is consistent with this hypothesis. If a past TDE also produced the disc and is responsible for the quiescent emission in \target, and the accretion rate is tracking the fallback rate, the lack of evolution implies that the TDE occurred at a time much earlier than the time at which we are currently detecting the QPEs. To see this, if we denote $L_0$ and $L_1$ as the luminosities at times $t_0$ and $t_1$, where $t_0$ is the time since disruption and $t_1 = t_0+\Delta t$ with $\Delta t = 3.5$ years (the time over which \target has been monitored), then it follows that
\begin{equation}
    t_0 = n \Delta t\left(1-L_1/L_0\right)^{-1}.
\end{equation}
Here we assumed that the luminosity tracks the fallback rate, where the latter scales as $t^{-n}$, with $n = 5/3$ if the object was completely destroyed \citep{phinney89} or $n = 9/4$ if it was partially destroyed \citep{coughlin19} and $t$ is time since disruption. Since $L_1/L_0 \simeq 1$ for \target, it follows that the star must have been destroyed well before $3.5$ years ago. 

Alternatively, it may be the case that the accretion rate onto the black hole is no longer tracking the fallback rate, but is instead evolving viscously \citep{cannizzo90}. While X-ray TDEs detected by transient surveys show declining luminosities and X-ray temperatures  with time \citep{2024ApJ...966..160G}, it seems likely that at sufficiently late times, the X-ray emission would evolve on the (in principle much longer) viscous timescale, rather than the fallback time of the debris \citep{auchettl17}. The ``plateau'' phase in the late-time optical/UV emission observed from some TDEs has been interpreted to arise from such a delay \citep{mummery}, and the fact that we are seeing no evolution in the quiescent X-ray flux from this system could imply that the same trend occurs at later times in the X-rays, consistent with theoretical models \citep{lodato11}.

\subsection{Implications for the model consisting of repeating partial tidal disruption of a white dwarf in a highly eccentric orbit}\label{sec:higheccTDE}
\citet{kingqpemodel} suggested that QPEs represent the repeated partial and tidal stripping of a white dwarf by a supermassive black hole (but of lower mass; see also \citealt{zalamea10}). In this model, the pericenter distance is highly relativistic: since a small amount of mass is removed from the star to power the accretion (making the standard assumption of the radiative efficiency of accretion; see Section \ref{sec:energetics}), the pericenter distance of the star is $r_{\rm p} \simeq 2 r_{\rm t} \simeq 2R_{\star}\left(M_{\bullet}/M_{\star}\right)^{1/3}$, and with\footnote{This assumes that the star is of low mass and the electrons are non-relativistic; the pericenter distance only becomes more relativistic as the white dwarf mass grows and its radius shrinks more rapidly.} $R_{\star} = 0.011 R_{\odot}\left(M_{\star}/(0.6 M_{\odot})\right)^{-1/3}$ \citep{nauenberg72}, $M_{\star} = 0.18 M_{\odot}$ \citep{king22}, and $M_{\bullet} = 2.3\times 10^{5}M_{\odot}$ \citep{king22}, $2 r_{\rm t} \simeq 7.3 GM_{\bullet}/c^2$. Even though the mass ratio is extreme, the timescale over which the period shrinks due to gravitational-wave emission is cosmologically short, which can be seen from the $e \simeq 1$ and $M_{\bullet} \gg M_{\star}$ limit of Equation 5.6 of \citet{peters64} when written in terms of the period of the orbiter and the pericenter distance of the orbit, which is 
\begin{equation}
    \dot{T}
    \simeq -\frac{85\pi}{4\sqrt{2}}\frac{M_{\star}}{M_{\bullet}}\frac{1}{x^{5/2}}\left(\frac{T}{T_{\rm p}}\right)^{2/3}, \label{Tdot}
\end{equation}
where we set $r_{\rm p} = x GM_{\bullet}/c^2$ and $T_{\rm p} = 2\pi r_{\rm p}^{3/2}/\sqrt{GM_{\bullet}}$. With the pericenter distance fixed -- which is a good approximation until the final stages of the inspiral; note that, from equations 5.5 and 5.6 of \citet{peters64}, $\dot{r}_{\rm p}/\dot{a} \propto \left(1-e\right)^2$ when $e \simeq 1$ and the mass ratio is small -- the gravitational-wave inspiral time that follows from the preceding equation is
\begin{equation}
\begin{split}
    &t_{\rm gw} \simeq 200\textrm{ yr} \\ 
    &\times\left(\frac{x}{10}\right)^{7/2}\left(\frac{M_{\bullet}}{10^5 M_{\odot}}\right)^{2/3}\left(\frac{M_{\bullet}/M_{\star}}{10^6}\right)\left(\frac{T_0}{\textrm{ 1 hr}}\right)^{1/3}.
\end{split}
\end{equation}
With $x = 7.3$, $M_{\bullet} = 2.3\times 10^{5}M_{\odot}$, and $M_{\star} = 0.18M_{\odot}$, this gives $t_{\rm gw} \simeq 200$ yr. 

\begin{figure*}
   \includegraphics[width=0.495\textwidth]{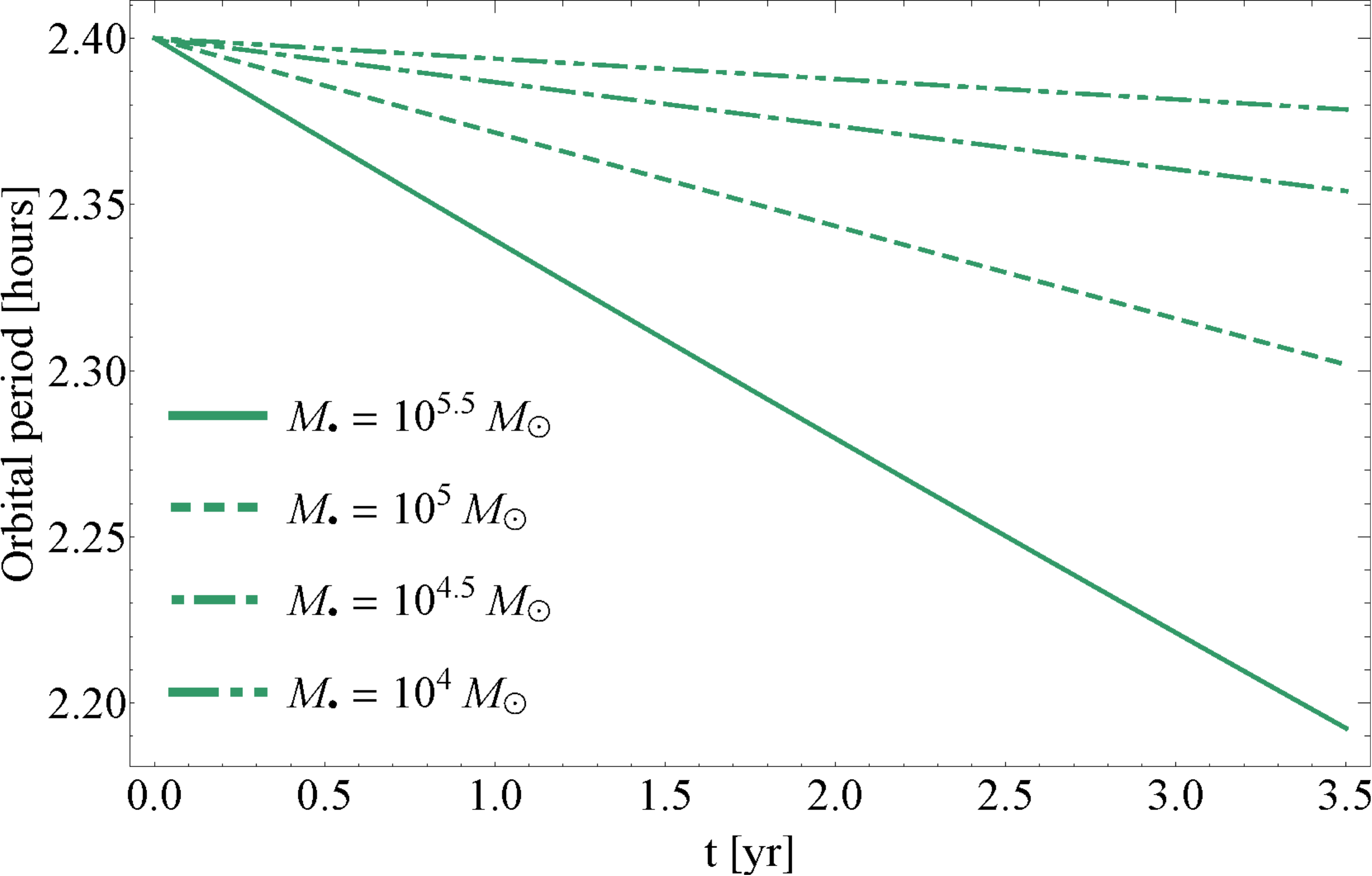}
        \includegraphics[width=0.495\textwidth]{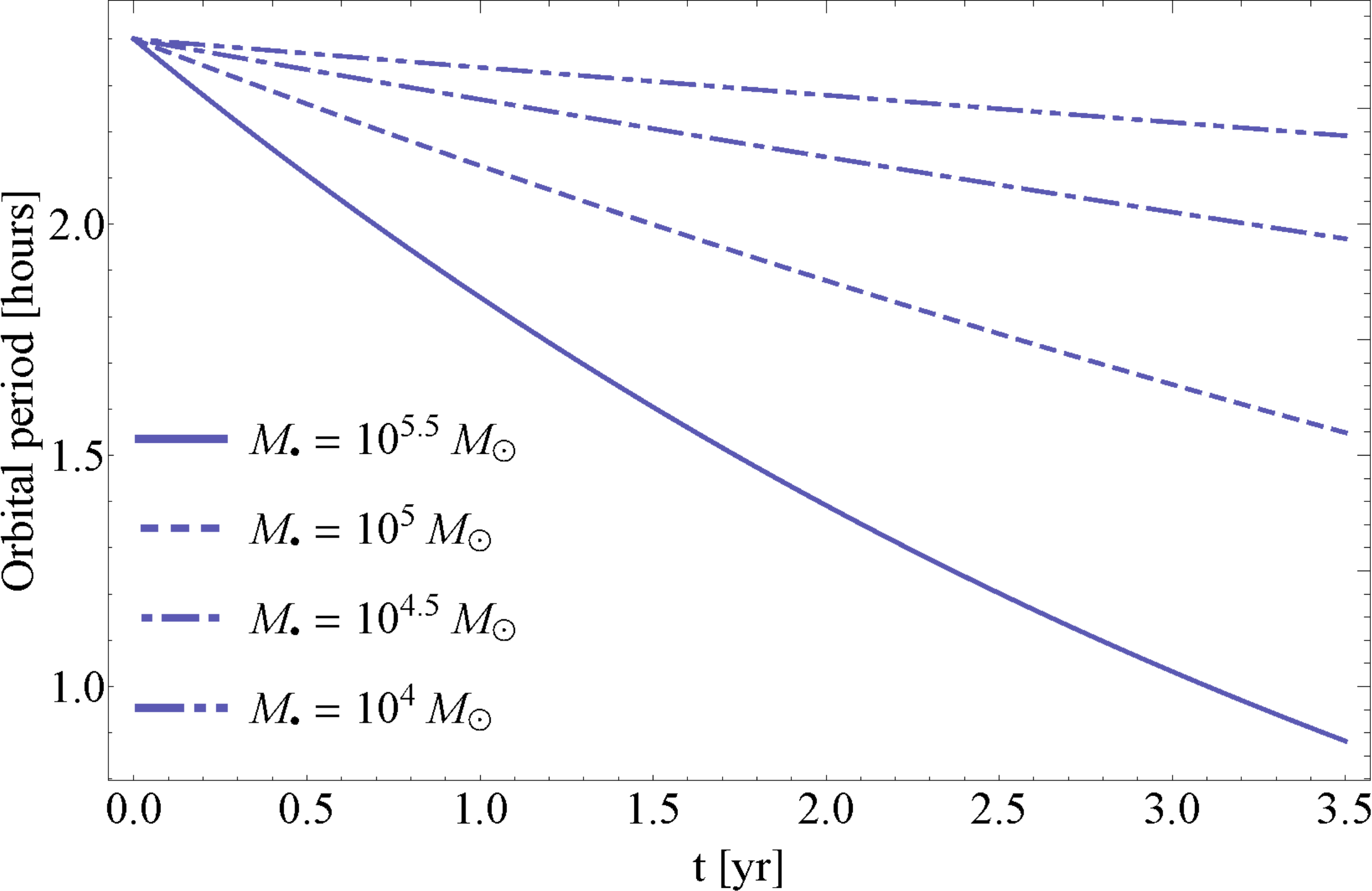}
    \caption{\textbf{The evolution of the orbital period due to gravitational-wave emission of a white dwarf around a massive black hole}. The black hole mass is shown in the legend and the stellar mass is $M_{\star} = 0.2 M_{\odot}$ (left) and $M_{\star} = 0.4 M_{\odot}$ (right). In each case we adopted a pericenter distance of $2r_{\rm t}$ (with the appropriate stellar radius, stellar mass, and black hole mass) and an initial orbital period of $2.4$ hours, which establishes the initial semimajor axis of the orbit and the initial orbital eccentricity for integrating the general (i.e., including eccentricity) \citet{peters64} equations. For low-mass white dwarfs and low-mass black holes, the orbital period does not decay by more than $\sim 0.1$ hours, which is consistent with observations (see Figure \ref{fig:fig2}). }
    \label{fig:gw_inspiral_peters}
\end{figure*}

Figure \ref{fig:gw_inspiral_peters} shows the evolution of the orbital period of the white dwarf using the \citet{peters64} equations for the decay in semimajor axis and eccentricity, where we set the initial period to $2.4$ hours and the pericenter distance to $2 r_{\rm t}$, which establish the initial semimajor axis and eccentricity. In the left (right) panel we adopted a stellar mass of $M_{\star} = 0.2 M_{\odot}$ ($M_{\star} = 0.4 M_{\odot}$), and the black hole mass is indicated in the legend; note that Equation \eqref{Tdot} predicts $\Delta T \simeq -0.085$ hr over 3 years for these parameters, $M_{\star} = 0.2 M_{\odot}$, and $M_{\bullet} = 10^5 M_{\odot}$, which agrees effectively exactly with the value of the dashed curve in the left panel at $t = 3$ yr. The observations of \target presented here suggest that the recurrence time of the flares may have declined from $\sim 2.4$ hours to $\sim 2.3$ hours from 2020 to 2022 (though there is no corresponding decline between 2022 and 2024), and if we attribute this change in period to gravitational-wave decay, then the parameters suggested in \citet{king22} -- namely a black hole mass of $2.3\times 10^5 M_{\odot}$ and a white dwarf mass of $0.18 M_{\odot}$ -- are broadly consistent with observations. On the other hand, a more massive star is strongly ruled out, unless the black hole is in the IMBH regime. 

\begin{sloppypar}

Since the (Galactic) white dwarf mass distribution is strongly peaked at $\sim 0.6 M_{\odot}$ \citep{obrien24} (and the production of a $\sim 0.2 M_{\odot}$ white dwarf would require mass exchange with a binary companion), this suggests that the original white dwarf had a mass closer to $\sim 0.6 M_{\odot}$ and was the core of a red giant, the envelope of which was stripped during the initial tidal interaction with the black hole -- as put forward by \citet{kingqpemodel}. For the same black hole mass, however, $2 r_{\rm t}$ for a $0.6 M_{\odot}$ white dwarf is $\sim 3.2 GM_{\bullet}/c^2$, implying that the black hole must be spinning and the orbit of the white dwarf must be prograde to avoid direct capture. Since the minimum pericenter distance an object can attain around a spinning black hole without being directly captured is \citep{will12, coughlin22} $r_{\rm min} = \left(1+\sqrt{1-a}\right)^2$, the black hole spin must satisfy $a \gtrsim 0.34$ to be able to partially tidally disrupt any white dwarf without direct capture, and the angular momentum of the orbit of that white dwarf would be exactly aligned with the black hole spin. For a black hole spin of $a = 0.998$, the number of encounters in the pinhole and full-loss cone regime -- from which the star must have originated because the tidal radius of the giant (i.e., the star on average) is orders of magnitude larger than that of the white dwarf -- that come within $2r_{\rm t}$ and are not directly captured is, with the methodology described in \citet{coughlin22}, $\simeq 5.5\%$, making such an encounter rare. It is also difficult to see how the mass transfer could be stable, given that the tidal radius of the white dwarf increases with decreasing stellar mass (characteristic of polytropic stars), and neither tides nor gravitational-wave emission modifies the pericenter distance significantly for such extreme-mass ratio systems (see the discussion in \citealt{cufari23, bandopadhyay24} relevant to TDEs of stars on bound orbits where the same arguments apply). 
\end{sloppypar}

\subsection{Implications for the low-eccentricity EMRI Hypothesis}\label{sec:shubham}
\begin{figure*}
    \centering
    \includegraphics[width=0.48\textwidth]{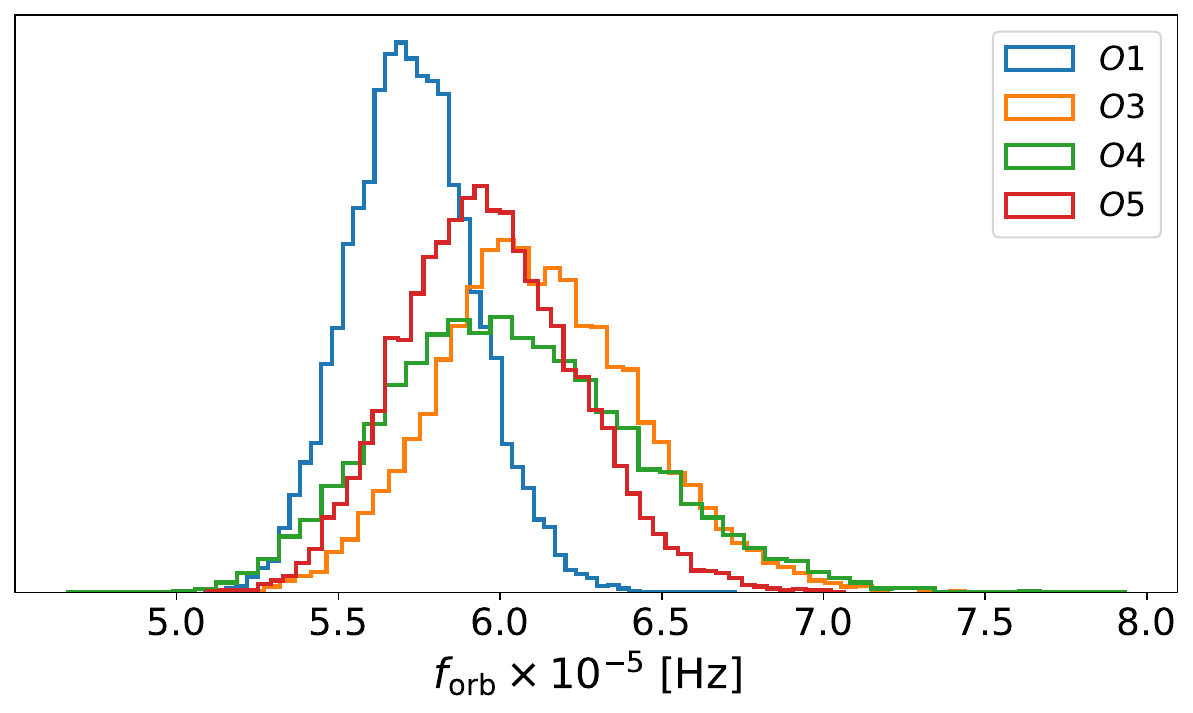}
        \includegraphics[width=0.48\textwidth]{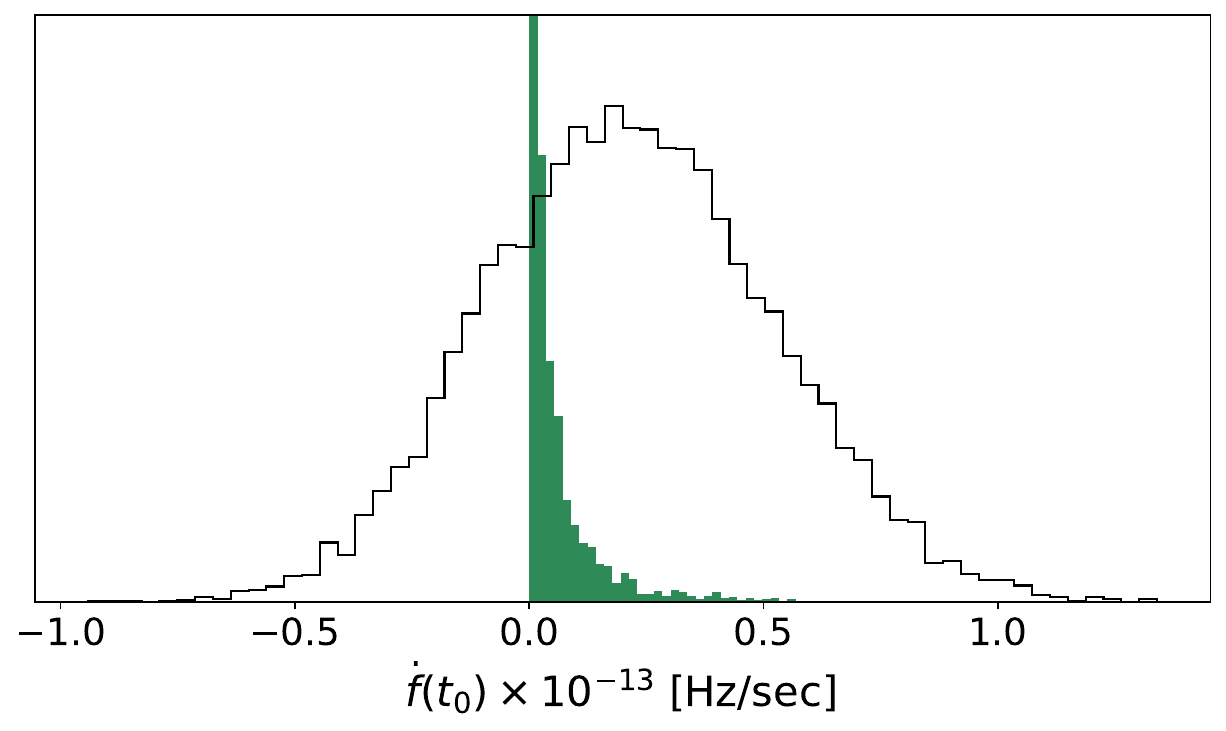}
    \caption{{\bf Left panel}: $10^4$ samples drawn from the posterior distribution of orbital frequencies $f_{\rm orb}$ inferred from the XMM data in run $\mathcal{O}n$ where n: 1, 3, 4 and 5 corresponding to XMM\#1,3,4 and 5. {\bf Right panel}: The posterior distributions (black solid line) from the data and the prior-predictive distribution (green filled bars) from the vacuum-EMRI model of $\dot{f}(t_0)$ where $t_0 = 2020$ for the XMM1 run. The vacuum EMRI predicts a slower evolution than inferred from the data.}
    \label{fig:forbsamples}
\end{figure*}

The long-term data for eRO-QPE2 can also be checked for consistency with the vacuum-EMRI hypothesis in low eccentricity configurations, as suggested by~\cite{Zhou:2024nuu}. Consider first the measured eruption times in all observation runs except XMM2\footnote{XMM2 only captures two full eruption events making it unsuitable for the analysis described in the text (which requires data from at least three peaks).}, denoted as $\hat{t}_{i}^n \pm \Delta{\hat{t}}_i^n$ for the $i^{\rm th}$ eruption within the $n^{\rm th}$ run with estimated $1\sigma$ errors $\Delta\hat{t}_i^n$ as described in Sec~\ref{sec: results}. The measured QPE period $\hat{T}_i^n$ can be estimated as $\hat{T}_i^n = \hat{t}_{i+1}^n - \hat{t}_{i}^n$ with errors $\Delta\hat{T}_i^n = \sqrt{(\Delta\hat{t}_{i+1}^n)^2 + (\Delta\hat{t}_{i}^n)^2}$. As shown in Fig.~\ref{fig:fig2}, $\hat{T}_i^n$ follows a long-short pattern, most clearly visible in the XMM1 run which, neglecting disk precession~\citep[see e.g.][]{alessiamodel,Arcodia:2024taw}, can be generically related to the Companion object's (CO's) orbital period as~\citep{Zhou:2024nuu} $\hat{T}_{\rm orb,i}^n = \hat{T}_i^n + \hat{T}_{i+1}^n$ with errors $\Delta\hat{T}_{\rm orb,i}^n = \sqrt{(\Delta\hat{T}_i^n)^2 + (\Delta\hat{T}_{i+1}^n)^2}$. Over a single run, we can treat the true long-timescale-averaged orbital period of the CO, $T_{\rm orb}^n \sim {\rm hours}$, as a constant since the dissipation timescales for the EMRI described by the priors below are much longer ($\sim$ years). With Gaussian likelihoods and flat priors on the $i^{\rm th}$ observation $T_{\rm orb,i}^n$, the posterior on $T_{\rm orb}^n$ following Bayes' theorem is given as 
$$p(T_{\rm orb}^n|\{\hat{T}_{\rm orb,i}^n\}) \propto \prod_i \mathcal{N}(\hat{T}_{\rm orb, i}^n|T_{\rm orb}^n, (\Delta\hat{T}_{\rm orb, i}^n)^2).$$ Samples $\tilde{T}_{\rm orb}^n$ drawn from the above posterior give corresponding posterior samples for the orbital frequency of the EMRI, $\tilde{f}_{\rm orb}^n = 1/\tilde{T}_{\rm orb}^n$, which are plotted in the left panel of Figure~\ref{fig:forbsamples} for various observation runs. Here, we note that a simple linear fit over the mean orbital periods in the different observation runs, $\langle \tilde{T}^n_{\rm orb,i} \rangle$ using the \texttt{linregress} module in \texttt{scipy.stats}~\citep{2020SciPy-NMeth} yields the slope $\dot{T}_{\rm orb} \approx -(1.4\pm1.1)\times 10^{-4}$ hours/day, such that $\dot{T}_{\rm orb}/2 \approx -(0.7\pm1.1)\times 10^{-4}$ hours/day which is consistent with our fit for the QPE period in Sec.~\ref{sec:linmix}. Furthermore, the time evolution of $f_{\rm orb}$ at $t_0 = 2020$ can be described to linear-order as $f(t) \approx f(t_0) + (t-t_0)\dot{f}_{\rm orb}(t_0)$. In a Bayesian predictive analysis, we can then check for consistency between observations and the EMRI hypothesis by comparing the observed value of $\dot{f}_{\rm orb}(t_0)$ with its prior-predictive distribution under the EMRI model\footnote{While higher-order derivatives may be required to better approximate the evolution of $f_{\rm orb}(t)$, they remain poorly constrained by current observations and are thus ignored in our analysis.}. 

To obtain the prior-predictive distribution, we evaluate the inspiral trajectories in generic low-eccentricity orbits around a Kerr black hole as described by the \texttt{5PNAAK} vacuum-EMRI model in the \texttt{FastEMRIWaveforms (FEW)} package~\citep{Katz:2021yft}. This model ignores perturbative effects from CO-disk interactions \citep[see e.g. ][]{Speri:2022upm, duqueeccentric} which however are small compared to the observational uncertainties on $\tilde{f}_{\rm orb}$ described above. We set the following conservative (uninformative) priors on the vacuum EMRI parameters describing the inspiral trajectory: the primary massive black hole (MBH) and CO masses follow log-uniform distributions, $M \sim \log\mathcal{U}[10^4,10^7]$, $M_* \sim M\times\log\mathcal{U}[10^{-5},10^{-4}]$, the dimensionless MBH spin, $a \sim \mathcal{U}[0.01,0.99]$, the orbit's initial eccentricity, $e(t_0) \sim \mathcal{U}[0.01,0.1]$, and the source's initial inclination with respect to the spin direction of the MBH, $\iota(t_0) \sim \mathcal{U}[0,\pi],$\footnote{$\iota(t_0) > \pi/2$ implies retrograde EMRI orbits.} all follow uniform distributions. The trajectories are initialized at the observed frequency samples during the first run, i.e. $f_{\rm orb}(t_0) = \tilde{f}_{\rm orb}^{\rm n=XMM1}$, and are evolved for the entire XMM temporal baseline.

The results of the prior-predictive analysis are presented in the right panel of Fig.~\ref{fig:forbsamples}. We find that, while the prior-predictive distribution is consistent with the posterior distribution of $\dot{f}_{\rm dot}(t_0)$, the average rate of evolution predicted by the vacuum EMRI model is $\approx$  an order-of-magnitude smaller than the observations. In other words, a vacuum EMRI in eRO-QPE2 would evolve significantly slower than the putative 0.1 hours over 3.5 years. Other models, such as drag-dominated EMRI inspirals~\citep{itaimodel2,Arcodia:2024taw}, or models of intermediate-mass-ratio inspirals~\citep{Amaro-Seoane:2018gbb}, or combinations thereof, can potentially explain the boosted rate of evolution, and should thus be investigated upon confirming the putative decline in \target.

Finally, a necessary condition for a vacuum-two-body GW inspiral is $\dot{f}_{\rm orb} > 0$, which is satisfied by $\approx 75 \%$ of samples of the data from the posterior distributions (right panel of Fig. \ref{fig:forbsamples}). Thus, the full range of vacuum-EMRI inspiral models are at most $75\%$ consistent with the 3.5-year eRO-QPE2 data. 

\section{Conclusions}
By performing timing and spectral analysis of \target's eruptions sampled five times over a period of 3.5 years we conclude:
\begin{itemize}
    \item The mean time between subsequent eruptions, i.e., the recurrence time, is constant between 2022 and 2024, with a hint of a decay of $\approx 0.1$ hr between August 2020 and June 2022\footnote{During the preparation of this paper, the authors of the preprint \citet{Arcodia:2024taw} also studied \target's long-term evolution, fitting individual eruptions with a Gaussian model (compared to our asymmetric Gaussian fits). These authors concluded that \target's recurrence times change from one observation to another which they attribute to either a gradual decline or evolution contaminated/dominated by modulations in arrivals of eruptions. This is distinct from our conclusion, and suggests that there maybe a fitting-function dependence to the trends that one infers from the data. If we consider the model-independent Bayesian blocks methodology, the evidence for a reduction in mean recurrence time is even less robust (see bottom panels of Fig. \ref{fig:fig2}).}. 
    \item The energy spectra of both the eruptions and the quiescence have remained stable over this 3.5 year period both in terms of shape and luminosity. This is consistent with \citet{Arcodia:2024taw}'s conclusions. 
    \item A low-mass ($\sim 0.2 M_{\odot}$) white dwarf partially tidally stripped by a $\sim 10^{5}M_{\odot}$ would experience a gravitational-wave-induced decay in its orbital period that is consistent with the \emph{average} $\sim 0.1$ hour reduction that is (tentatively) observed, but the detailed evolution of the recurrence time, and specifically the lack of period evolution over the two years from 2022 to 2024 (see Figure \ref{fig:fig2}), is not consistent with the monotonic decline that is expected from gravitational-wave emission.
    \item Finally, we find that observed stability is consistent with a vacuum EMRI scenario for  a wide range of parameters (EMRI mass ratio, eccentricity, spin, inclination). In fact, a vacuum EMRI predicts almost an order of magnitude slower evolution that the putative 0.1 hrs over 3.5 years. This elucidates the need for drag-dominated, the so-called ``dirty'' EMRI frameworks, to accurately model these systems.  
\end{itemize}

\section*{Acknowledgments}
E.R.C. acknowledges support from NASA through the \emph{Neil Gehrels Swift} Guest Investigator Program, proposal number 1922148. E.R.C.~acknowledges additional support from the National Science Foundation through grant AST-2006684, and from NASA through the Astrophysics Theory Program, grant 80NSSC24K0897. This research was supported in part by grant NSF PHY-2309135 to the Kavli Institute for Theoretical Physics (KITP). M.Z. acknowledges support from the Czech Science Foundation through grant GA\v{C}R Junior Star No. GM24-10599M. Support for this work was provided by NASA through
grant GO-17447 from the Space Telescope Science Institute,
which is operated by AURA, Inc., under NASA contract
NAS5-26555. D.R.P would like to thank Muryel Guolo for feedback on the paper. 

\section*{Data Availability:}
The data to reproduce Figures 1, 2 and 3 can be found on {\tt zenodo:} \url{https://zenodo.org/records/11415786}



\clearpage

\newpage

\newpage
\bibliographystyle{aasjournal}
\bibliography{arxiv}

\end{document}